\documentclass[12pt]{article} 

\usepackage[margin=1in]{geometry}
\usepackage[utf8]{inputenc}
\usepackage{graphicx}

\usepackage{amsmath}
\usepackage{amssymb}
\usepackage{newtxmath}
\DeclareMathAlphabet{\mathpzc}{T1}{pzc}{m}{it}
\def\given{\mid}
\usepackage{bm}

\usepackage{caption}
\usepackage[style=nature,
					backend=biber,
					sortcites=true,
]{biblatex}

\usepackage[section]{placeins}
\AtEveryBibitem{%
 \clearfield{url}%
   \clearfield{month}%
 \clearfield{issn}%
 \clearfield{doi}
}

\usepackage{times}

\usepackage{url}

\usepackage{changepage}

\usepackage{setspace}
\usepackage[font=small,labelfont=bf]{caption}

\usepackage[usenames,dvipsnames,svgnames,table]{xcolor}

\def\infinity{\rotatebox{90}{8}}

\usepackage{ marvosym }

\usepackage{authblk}

\title{COVID-19 confines recreational gatherings in Seoul to familiar, less crowded, and neighboring urban areas}

\author[1,2,3]{Jisung Yoon}
\author[3,4]{Woo-Sung Jung}
\author[5,*]{Hyunuk Kim} 

\affil[1]{Kellogg School of Management at Northwestern University, Evanston, IL 60208, USA.}
\affil[2]{Northwestern Institute on Complex Systems, Evanston, IL 60208, USA.}
\affil[3]{Department of Industrial and Management Engineering, Pohang University of Science and Technology, Pohang 37673, Republic of Korea}
\affil[4]{Department of Physics, Pohang University of Science and Technology, Pohang 37673, Republic of Korea}
\affil[5]{Department of Administrative Sciences, Metropolitan College, Boston University, MA 02215, USA}
\affil[*]{Corresponding author: uk@bu.edu}

\date{\today}

\begin{document}

\baselineskip24pt

\maketitle

\section*{Abstract}

Recreational gatherings are sources of the spread of infectious diseases. Understanding the dynamics of recreational gatherings is essential to building effective public health policies but challenging as the interaction between people and recreational places is complex. Recreational activities are concentrated in a set of urban areas and establish a recreational hierarchy. In this hierarchy, higher-level regions attract more people than lower-level regions for recreational purposes. Here, using customers' motel booking records which are highly associated with recreational activities in Korea, we identify that recreational hierarchy, geographical distance, and attachment to a location are crucial factors of recreational gatherings in Seoul, Republic of Korea. Our analyses show that after the COVID-19 outbreak, people are more likely to visit familiar recreational places, avoid the highest level of the recreational hierarchy, and travel close distances. Interestingly, the recreational visitations were reduced not only in the highest but also in low-level regions. Urban areas at low levels of the recreational hierarchy were more severely affected by COVID-19 than urban areas at high and middle levels of the recreational hierarchy.


\section*{Introduction}

Human urban activities are principal elements of social phenomena, including the growth of cities~\parencite{verbavatz2020growth, bettencourt2020demography}, economies~\cite{storz2015mobility, park2019global}, and epidemics~\cite{chang2021mobility}.
They tend to be concentrated in parts of cities and form a hierarchy of geographical areas, where regions at upper levels attract more people than those at lower levels~\cite{barthelemy2016structure, batty2013new, pan2013urban}. A person may frequently visit a popular region, often referred to as a \textit{hotspot}~\cite{roth2011structure, bassolas2019hierarchical, louail2015uncovering}, even though it is far from living areas. 

Strong urban hierarchy raises various concerns during a pandemic~\cite{ahmed2020inequality, yu2021racial, van2020covid}. The spread of infectious diseases would be broad and prevalent if it originates from a hotspot at the top of the hierarchy~\cite{albert2000error,pastor2003epidemics, kang2020coronavirus, gould1994spatial}. The economic impact of a pandemic also differs by hierarchy level. The income of the populations working in the informal economy, which is usually located at low hierarchy levels, was negatively affected by the COVID-19 pandemic~\cite{narula2020policy}. Despite its importance to human activities, the urban hierarchy has been rarely considered when analyzing behavioral changes in response to a pandemic~\cite{song2010modelling, schlapfer2021universal, batty2021london, nouvellet2021reduction,moro2021mobility}.

Here, by using individual-level motel booking records (see Methods) from a leading Korean accommodation platform, we compare a visitation pattern before and after the COVID-19 outbreak. According to a market report in 2021~\cite{accomodationstat2021}, 32.6\% of the platform's mobile application installers were in their 20s, 35.4\% were in their 30s, 23.9\% were in their 40s, and 6.4\% were above 50s. Additionally, 37.7\% were females, and 63.3\% were males. Therefore, low- and middle-income populations are likely to be the primary users of the platform and the Korean accommodation market.

Motels are often located near recreational places such as pubs, nightclubs, restaurants, and cafes in urban hotspots~\cite{lashley2016routledge}.  Especially, in recent years that our data cover, the Korean motel industry has transformed itself into an entertainment industry that provides physical spaces for relaxation and cultural activities~\cite{kim2018qualitative}. 
Young Koreans increasingly book motels for intimate relationships and partying with friends because motels are more affordable and accessible than hotels~\cite{Hu2016}. For these reasons, we use our motel reservation data as a proxy for recreational gatherings in Seoul. The hierarchy of recreational urban areas which is extracted from our motel booking data is referred to as the \textit{recreational hierarchy}. Our analyses show that recreational hierarchy, geographical distance, and attachment to a location are important factors of recreational gatherings in Seoul, the largest city of the Republic of Korea.

\section*{Results}

\begin{figure}[ht]
    \centering
    \includegraphics[width=\textwidth]{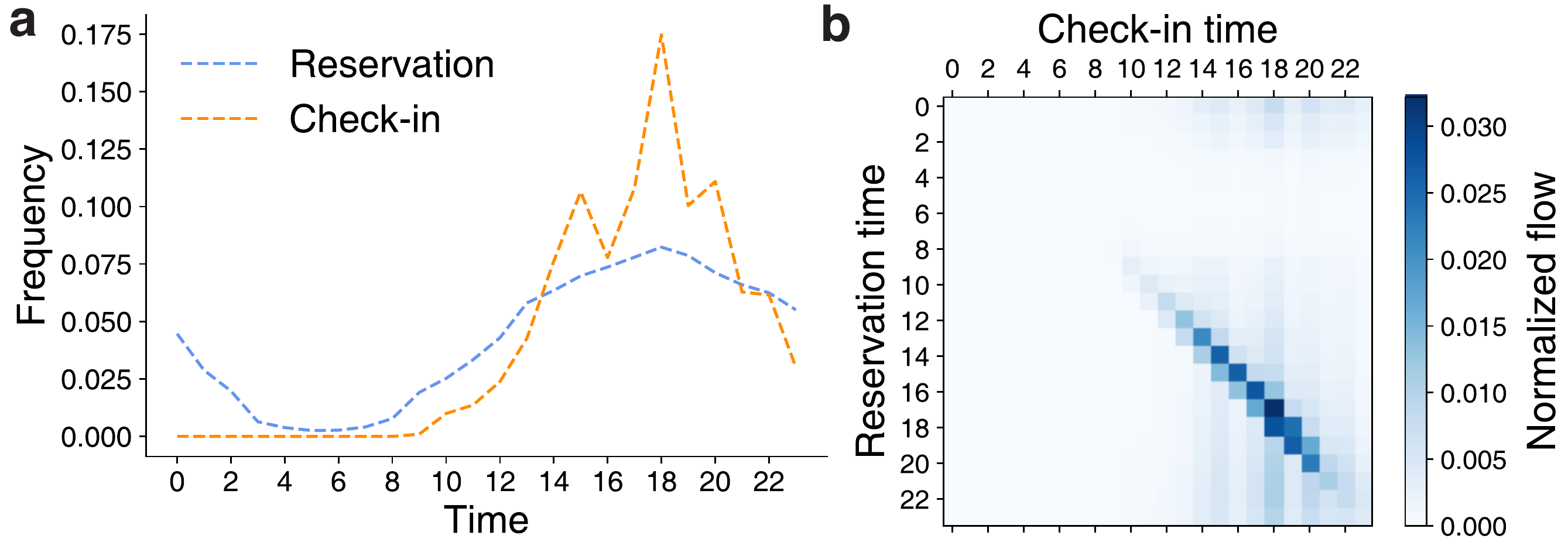}
    \caption{\textbf{(a) Reservation and check-in time distributions.}
 Reservations and check-in take place primarily after 12pm.
 \textbf{(b) A heatmap for the reservation and check-in times.}
 Both reservation and check-in times show a similar pattern and are concentrated after 12pm.}
    \label{fig:check_in_resrv}
\end{figure}

\begin{figure}
\centering
\includegraphics[width=\linewidth]{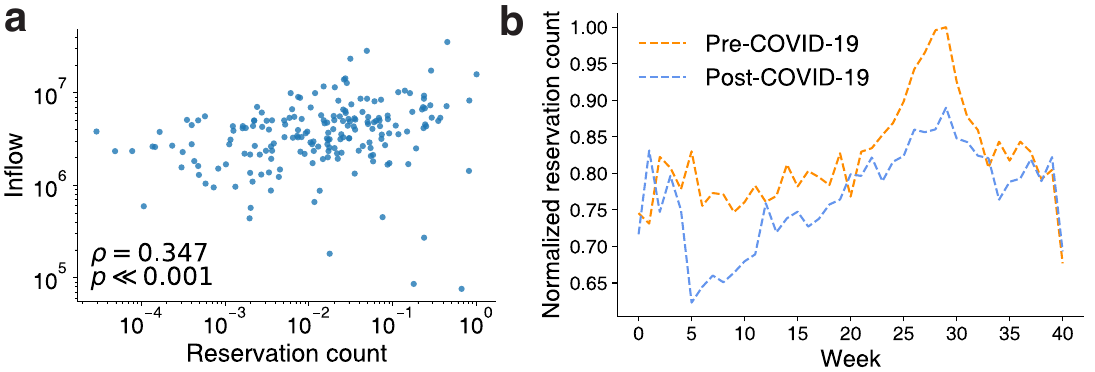}
\caption{\textbf{(a) A comparison of the reservation counts to the mobility inflows in Seoul, Republic of Korea.} We aggregate the mobility inflows at the level of the administrative division. Each dot represents an administrative division. A significant correlation supports that the reservation history data can be a good proxy for urban recreational gatherings in Seoul. \textbf{(b) Weekly reservation counts.} For a data privacy concern, we normalize the weekly reservation counts by the maximum weekly reservation count.}
\label{fig:validation_desc_fig}
\end{figure}

In our data, motel reservation and check-in times are concentrated after 12pm (Fig.~\ref{fig:check_in_resrv}a) and there is little time difference between reservation and check-in times (Fig.~\ref{fig:check_in_resrv}b). To validate whether our data capture recreational gatherings in Seoul to some extent, we compare reservation counts at the administrative division level with the mobility inflows from Seoul mobility data aggregating GPS locations (See Methods). The rank correlation between the reservation counts and the mobility inflows is significant ($\rho=0.347$, $p$-value $<0.001$; Fig.~\ref{fig:validation_desc_fig}a). The correlation becomes stronger if we only consider nighttime inflows (from 9 pm to 6 am, $\rho=0.400$, $p$-value $<0.001$).

To understand the effect of COVID-19 on recreational gathering behaviors, we split the data into two periods: \texttt{pre-COVID-19} (From January 21, 2019 to November 3, 2019) and \texttt{post-COVID-19} (From January 20, 2020 to November 1, 2020). January 20, 2020 is the first day that a COVID-19 infection case was reported in Korea. Both periods start from the fourth week of January and span 286 days. The weekly trend of reservation counts is shown in Fig.~\ref{fig:validation_desc_fig}b. A significant drop appears near Week 5, the first week of the official social distancing in the Republic of Korea. The reservation counts were recovered gradually to the normal state even after several restrictions were imposed.

\subsection*{Recreational hierarchy of Seoul}

\begin{figure}
\centering
\includegraphics[width=\linewidth]{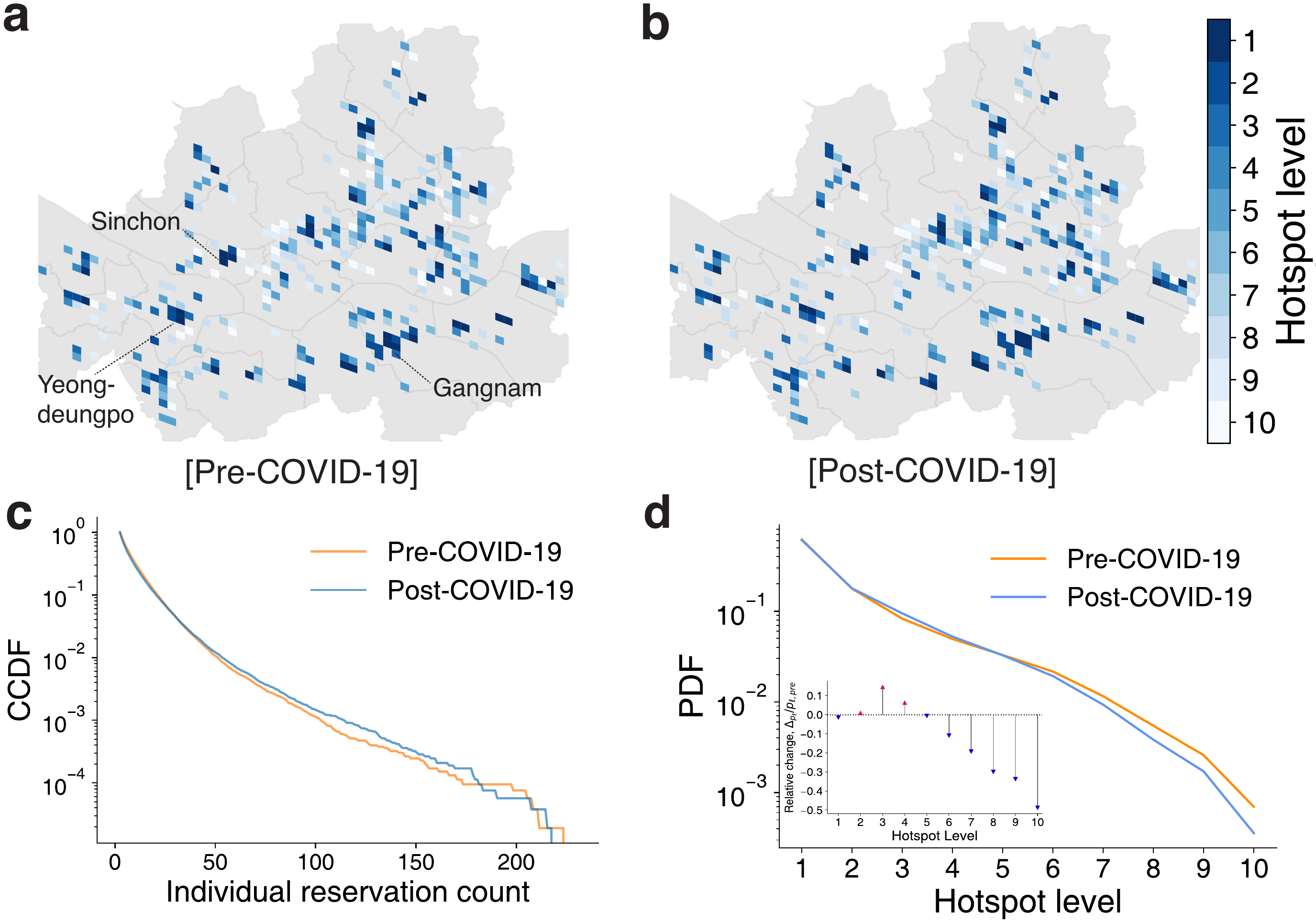}
\caption{ \textbf{(a-b) The pre- and post-COVID-19 recreational hierarchy maps.} Each cell represents a level-14 Google S2 cell colored by its hotspot level. The areas with grey color represent cells with no accommodations. \textbf{(c) The complementary cumulative distribution function (CCDF) of individual reservation counts.} On average, individual reservation counts decrease after the COVID-19 outbreak. \textbf{(d) The reservation count distribution by the hotspot level $p_{\ell}$.} The inset shows the relative change of $p_{\ell}, (p_{\ell, post} - p_{\ell, pre})/p_{\ell, pre} $. A red up-arrow indicates an increase of the probability and a blue down-arrow indicates a decrease of the probability compare to the distribution for the pre-COVID-19 period.
}
\label{fig:desc_fig}
\end{figure}

We assign each motel to a Google S2 cell (\url{https://github.com/google/s2geometry}). S2 cells are space tessellations that divide the Earth into cells of a similar size area. It is known as a robust, flexible spherical geometry \cite{googles2, wu2018travel, bassolas2019hierarchical}. We used level-14 S2 cells of which size ranges from 0.19$km^2$ to 0.40$km^2$ (on average 0.32$km^2$). Then, we aggregate the reservation counts by S2 cell and identify a hierarchy of cells by assigning a \textit{hotspot level}, an inverse decile rank of aggregated reservation counts, to each cell. Level 1 is the highest, and level 10 is the lowest level. Fig.~\ref{fig:desc_fig}a and Fig.~\ref{fig:desc_fig}b show the recreational hierarchy maps for both periods. The assigned hotspot levels are almost consistent for both periods. Cells with high levels correspond to popular recreational areas in Seoul such as Gangnam, Sinchon, and Yeongdeungpo Time Square (highlighted in Fig.~\ref{fig:desc_fig}a).

To further analyze behavioral changes induced by COVID-19, we take a subset of customers who have at least two reservation records in both the pre- and post-COVID-19 periods as the focus group. This focus group covers 30\% of the total customers in the pre-COVID-19 period and 26\% in the post-COVID-19 period. The distributions of individual reservation counts are similar for both periods, but the average individual reservation counts decreased after the COVID-19 outbreak (Fig.~\ref{fig:desc_fig}c; $\langle l_{pre} \rangle=9.200 > \langle l_{post} \rangle=8.757$; $\text{paired t-statistic}=8.820, \text{p-value} \ll 0.001$).

As shown in Fig.~\ref{fig:desc_fig}d, the majority of reservations (61.6\% for the pre-COVID-19 period, 60.7\% for the post-COVID-19 period) is concentrated in the top 10\% cells, while the bottom 10\% cells only have a few reservations (0.6\% for the pre-COVID-19 period, 0.3\% for the post-COVID-19 period), suggesting the inequality of recreational visitations on urban areas for both periods. Interestingly, COVID-19 affects the inequality of urban areas differently by the hierarchy level. The proportion of the highest level decreases after the COVID-19 outbreak (Fig.~\ref{fig:desc_fig}d inset). However, this proportion was not equally distributed across other levels. People visited levels 2, 3, and 4 rather than low levels ($l \leq 6$). Our findings show that the COVID-19 pandemic worsened the inequality across urban areas, in line with previous studies on income levels~\cite{belot2021unequal,zhang2022unequal} and costs of shutdown ~\cite{hevia2022covid}. We explain the worsening inequality by decomposing individual recreational gathering behaviors in the next section.

\subsection*{Factors of recreational gatherings}

\begin{figure}
\centering
\includegraphics[width=\linewidth]{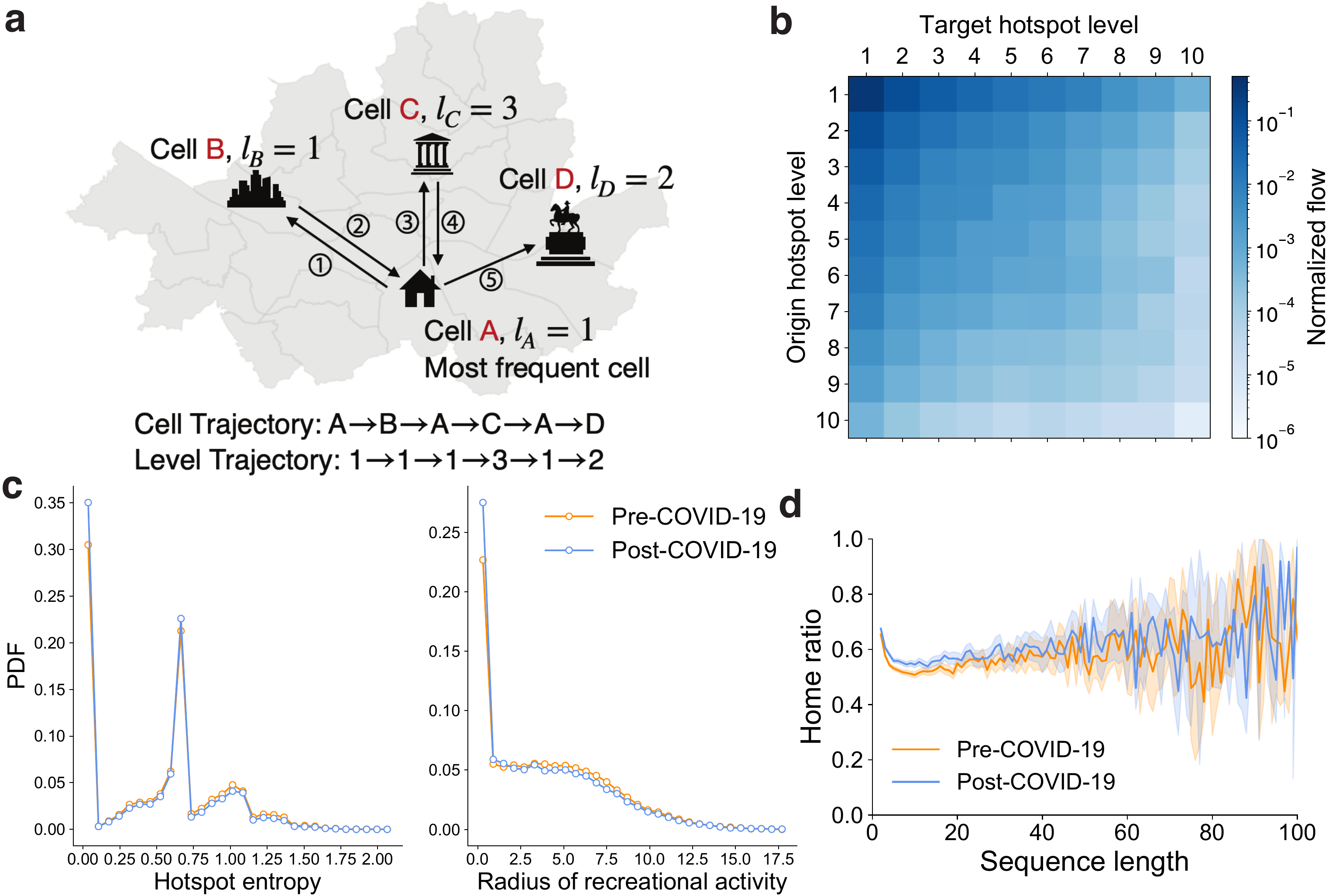}
\caption{\textbf{(a) An illustrative example of the cell and level trajectories.} \textbf{(b) The flow matrix $T^{data}$ for the pre-COVID-19 period.} The trips within the same cell are excluded. We also provide the flow matrix for the post-COVID-19 period in Supplementary Information (Fig.~S1). \textbf{(c) The distributions of the hotspot entropy $p_h^{data}$ and the radius of recreational activities $p_r^{data}$ for both periods.} After the outbreak, people explored less across the recreational hierarchy. \textbf{(d) Home ratios by sequence length for both periods.} Attachment to a location appears regardless of sequence length, implying the share of time in the recreational home remains constant after the outbreak.}
\label{fig:data_fig}
\end{figure}

Individual records can be converted to sequences of cells and hotspot levels. The arrows in Fig.~\ref{fig:data_fig}a represent a synthetic journey that consists of urban areas. The \textit{cell trajectory} of this example is $A \rightarrow B \rightarrow A \rightarrow C \rightarrow A \rightarrow D$. Note that the same place can appear multiple times. Based on the assigned levels of the cells, the \textit{level trajectory} is $1 \rightarrow 1 \rightarrow 1 \rightarrow 3 \rightarrow 1 \rightarrow 2$. For each trajectory constructed from the data, we define the most frequent cell as the \textit{recreational home}, so $A$ is the home in the example.

\subsubsection*{Hierarchy: Transition between levels}
We construct a flow matrix $T^{data}$ where $T^{data}_{ij}$ is the number of trips between level $i$ and $j$ normalized by the total flows (Fig.~\ref{fig:data_fig}b). We here exclude self-transitions, trips within the same cell, to focus on the transitions between different cells. The majority of transitions are concentrated in high levels of the hierarchy, and the flow matrix is almost symmetric. To check whether the transition from level $i$ to level $j$, $p(j \given i)$, depends on level $j$, we build a null model \cite{bassolas2019hierarchical} that $p(j \given i)$ is proportional to the total inflows to destination's level $j$ as follow,

\begin{equation}
T_{ij}^{null} = \sum_{k=1}^L T_{ik} \frac{\sum_{m=1}^L T_{mj}}{\sum_{m,k=1}^L T_{mk}},
\end{equation}

where $\sum_{k=1}^L T_{ik}$ is the total outflow from level $i$ and $ \sum_{m=1}^L T_{mj}/\sum_{m,k=1}^L T_{mk}$ is the fraction of the inflows to level $j$ (Supplementary Fig.~S2). Comparing the ratio of $T^{data}$ to $T^{null}$ (Supplementary Fig.~S3), we confirm that $T^{data}$ is close to $T^{null}$ at high levels of the hierarchy, while the ratios of the transitions from or to low levels of the hierarchy increase. Most of the transitions are at high levels of the hierarchy, and inflow and outflow are symmetric (Supplementary Fig.~S4; $R^2$=0.99 for both periods). Hence, transitions between hotspot levels are approximately independent of the previous place’s hotspot level $p(j \given i) \simeq p(j)$ which follows the reservation count distribution by the hotspot level $p_{\ell}$.

\subsubsection*{Hierarchy: hotspot entropy}
To measure the extent to which hotspot levels are diverse in individual records, we introduce the \textit{hotspot entropy}, $h$ (See Methods). $h$ is zero if the trajectory consists of cells of the same hotspot level. The maximum value of $h$ is $\ln {\text{(number of hotspot levels)}} = \ln{10} $. The distributions of $h$ for both periods are different ($\text{KS-statistic}=0.046, \text{p-value} \ll 0.001$) and shown in Fig.~\ref{fig:data_fig}c (left). Before the COVID-19 outbreak, 30\% of people stay only at a single level on average, while this proportion increases after the outbreak. Also, the mean hotspot entropy decreases ($\langle h_{pre} \rangle=0.518 > \langle h_{post} \rangle=0.475$), implying people are less likely to visit different levels.

\subsubsection*{Geographical distance}
The \textit{radius of recreational activities}, $r$ (See Methods), a variance of geographical distances from the recreational home of a sequence, quantifies how far the places in a trajectory are. The unit of $r$ is a kilometer (km). The distributions of $r$ show that the majority of people stays within a single cell without moving to other cells (Fig.~\ref{fig:data_fig}c, right). The likelihood of visiting distant places is inversely proportional to geographical distance, while the hierarchy leads people to move farther than expected (Supplementary Fig.~S5). Considering the radius of the biggest district in Seoul (Seocho district) that is about 5.523km, we can say that more than 33\% of the platform users in the pre-COVID-19 period visit places outside the home cell ($\sim$30\% for the post-COVID-19 period). Overall, $r$ decreases after the COVID-19 outbreak ($\langle r_{pre} \rangle=4.094 > \langle r_{post} \rangle=3.738$), and the distributions for both periods are significantly different  ($\text{KS-statistic}=0.056, \text{p-value}\ll0.001$). This evidence suggests that people tend to stay close to their recreational homes after the outbreak.

\subsubsection*{Attachment to a location}

Attachment to a location is an indicator of customer satisfaction and an important factor for the accommodation business~\cite{bowen2001relationship, kandampully2000customer}. In Fig.~\ref{fig:data_fig}d, we show the home ratio which is the fraction of the most frequent cell in a sequence. Interestingly, the home ratio is about 0.6 regardless of sequence length. The home ratio slightly increases after the COVID-19 outbreak for the light users who booked motels no more than 20 times, while there is no difference in the home ratio between the two periods for the heavy users who booked motels more than 20 times (Top 10\% users by sequence length).

\subsection*{A model for replicating reservation records}

\begin{figure}[ht]
\centering
\includegraphics[width= \linewidth]{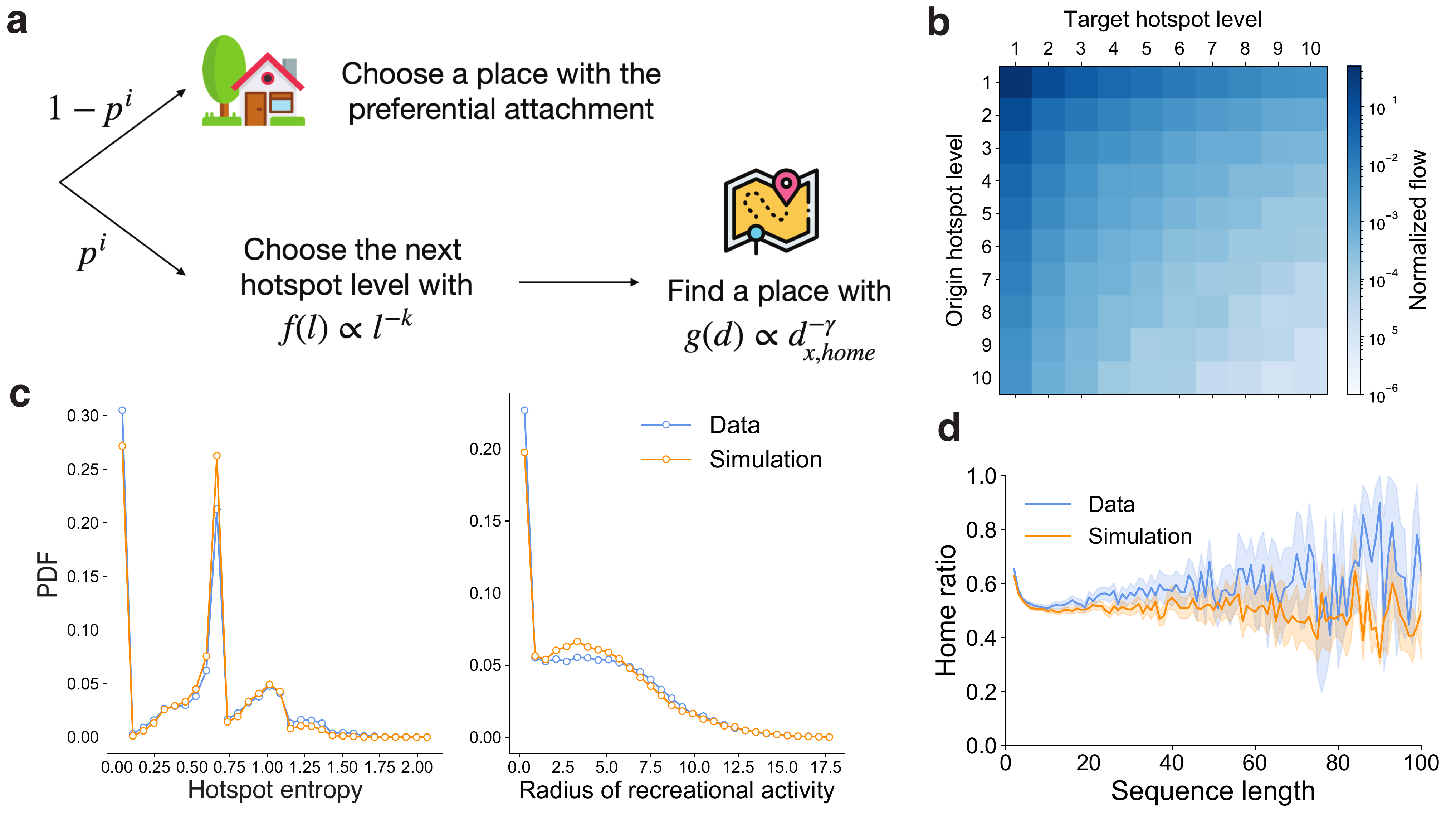}
\caption{\textbf{(a) A schematic diagram of our model.} $k$, $\gamma$, and $p$ control the effects of recreational hierarchy, geographical distance, and attachment to a location, respectively. $i$ is the index of an iteration. \textbf{(b) The flow matrix of the best model} \textbf{(c) The distributions of the hotspot entropy distribution $p_h$ (left) and the radius of recreational activities $p_r$ (right).} Blue lines are the empirical distributions, and orange lines are the simulation results. \textbf{(d) Home ratios by sequence length.} Here, we show the results from the best model for the pre-COVID-19 period. The best model for the post-COVID-19 period is in Supplementary Fig.~S6.}
\label{fig:model_result_fig}
\end{figure}

Our empirical analysis reveals that recreational hierarchy, geographical distance, and attachment to a location need to be considered simultaneously to explain recreational gatherings in Seoul, Republic of Korea. Leveraging our key findings on the individual movements, we develop a model reproducing their patterns and detecting behavioral changes during the COVID-19 pandemic (Fig.~\ref{fig:model_result_fig}a). Our model is motivated by the literature analyzing human mobility~\cite{song2010modelling, schlapfer2021universal, moro2021mobility}. First, an agent starts from an initial cell randomly picked from the cell-level reservation count distribution. Then, the agent explores places with probability $p^i$ or chooses a previously visited place in proportion to the frequency in the reservation history with probability $1-p^i$, where $p \in [0,1]$ controls the likelihood that the agent decides to explore places and $i$ is the number of iterations starting from one. As the iteration $i$ increases, the agent is more likely to choose previously visited places.

If the agent decides to explore places, the agent first chooses the hotspot level for the next move from the distribution $f(\ell) \propto \ell^{-k}$ ($k \in [0,\infinity)$). $k$ can be interpreted as the preference to high hierarchy levels of the agent. For instance, if $k=0$, the agent randomly selects the next level without considering recreational  hierarchy. On the other hand, with a large $k$, most reservations are concentrated in high hierarchy levels. Next, the agent chooses a place of the given level in proportion to the inverse of geographical distance with an exponent $\gamma \in [0,\infinity)$ which controls the likelihood of visiting distant places from the recreational home. If $\gamma=0$, the agent ignores the geographical distance and randomly picks the place with the given level.
In this step, the agent can pick the recreational home by imputing the $d_{home, home}=1$, and the recreational home cell can change according to the current history of the agent as the iteration proceeds. The steps above are repeated until we have a sequence of which length is equal to the length of the original trajectory. We keep the sequence length distribution from the data (Fig.~\ref{fig:desc_fig}c) to control the effect of sequence length. 

Through a grid search, we estimate the model parameters that minimize the Jensen-Shannon divergence (JSD) of the hotspot entropy distribution $p_h$, the reservation count distribution $p_l$, and the radius of recreational activities distribution $p_r$ between synthetic sequences and the data. Note that $p$ and $k$ affect $p_l$ and $p_h$, while $\gamma$ is independent of $p_l$ and $p_h$. Taking advantage of this property, we jointly optimize the model by searching the best $p$ and $k$ that minimize the product of JSD of $p_l$ and $p_h$, namely $JSD_{Entropy}$ and $JSD_{Hierarchy}$. Next, with the best $p$ and $k$, we fit the best $\gamma$ that minimizes $JSD_{Radius}$, JSD of $p_r$. For the grid search, we explore $p$ with dividing 0 to 1 into 51 bins (bin width = 0.02), $k$ with dividing 0 to 3 into 121 bins (bin width = 0.025), and $\gamma$ with dividing 0 to 5 into 201 bins (bin width = 0.025). We repeat the simulation ten times and average the estimated model parameters. 

Our model successfully reconstructs the flow matrix $T_{model}$, all distributions, and the retention of attachment to a location (Fig.~\ref{fig:model_result_fig}b-d). Fig.~\ref{fig:model_result_fig}b shows the flow matrix from the model, $T^{model}$. A matrix distance between $T^{model}$ and $T^{data}$ measured by the Frobenius norm $d_T = ||T^{model} - T^{data}||_F$ is 0.03, indicating the model reproduces the flows across the recreational  hierarchy. Although there are gaps in the first bin of the generated distributions and the home ratio, our model captures the overall patterns of individual movements well. It is difficult to model outliers who rarely move to other places. Furthermore, we compare the simulated reservation counts from the model and the actual reservation counts and find that the model also explains the reservation count well for both the pre- and post-COVID-19 periods (Supplementary Fig.~S7).

To better understand the effects of these factors on recreational gatherings, we examine the variant models that exclude each factor (Supplementary Fig.~S8). From the estimated $p,k$, and $\gamma$ for the best model, we construct the variant models by changing the target parameter while keeping other parameters the same. For the model without recreational  hierarchy, we simulate the model with $k=0$. In this model, an agent does not consider recreational  hierarchy, and this change results in a collapse of the model in the flow matrix ($d_T=0.35$; Supplementary Fig.~S8a) and $p_h$ (Supplementary Fig.~S8b). Similarly, we try the model without geography with $\gamma=0$ where an agent does not take into account geographical distance. This model still produces the comparable result on the hierarchy ($d_T=0.06$) and $p_h$, while $p_r$ is totally collapsed as expected (Supplementary Fig.~S8e). Lastly, we build the model without attachment to a location with $i=1$. In this model, the likelihood that an agent explores a place is always $p$ so that the likelihood does not depend on the iteration. The model without the attachment reproduces the hierarchy ($d_T=0.04$), weakly collapses in $p_h$ and $p_r$, but does not reproduce the retention of attachment to a location (Supplementary Fig.~S8i). In short, recreational hierarchy, geographical distance, and attachment to a location are indispensable factors of urban recreational gatherings in Seoul.

\begin{figure}[ht]
\centering
\includegraphics[width=\linewidth]{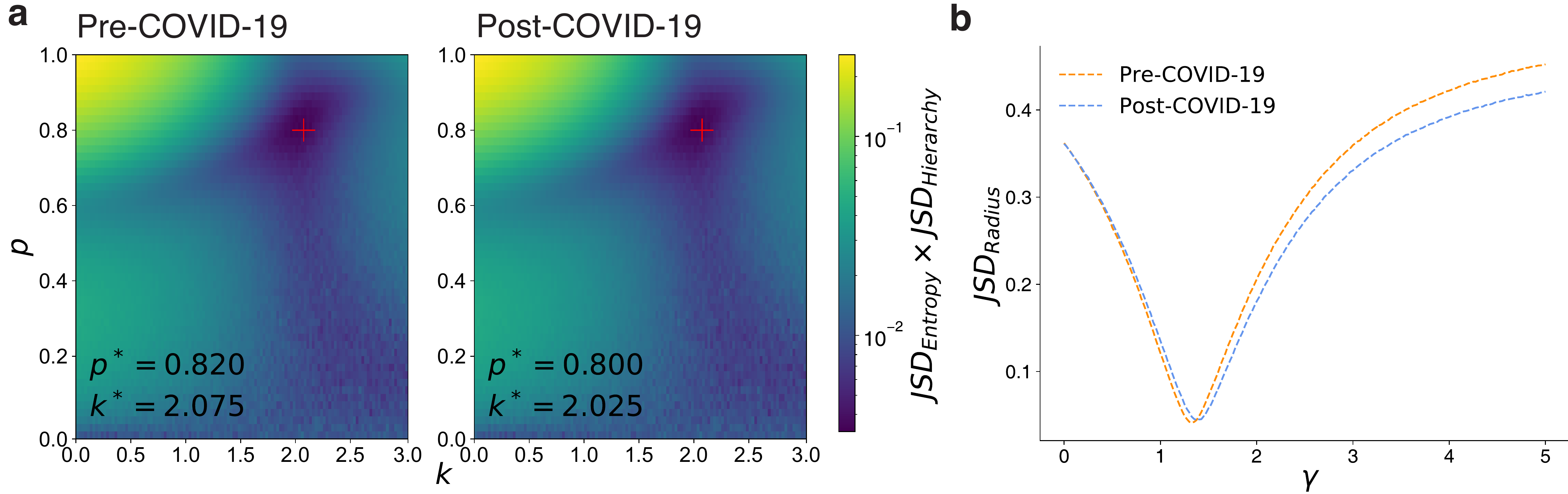}
\caption{
\textbf{(a) $JSD_{Entropy} \times JSD_{Hierarchy}$ varied by the model parameters $p$ and $k$}. Bottom annotated $p^{*}$ and $k^{*}$ are the best parameters for each period. \textbf{(b) $JSD_{Radius}$ varied by the model parameter $\gamma$.} We change $\gamma$ while keeping the best $p$ and $k$ from Fig.~\ref{fig:compare_fig}a.}
\label{fig:compare_fig}
\end{figure}

\subsection*{An external shock influences recreational gatherings}

We explore the influence of the COVID-19 outbreak on individual movements for recreational activities by comparing the best model result for each period. We show the $JSD_{Entropy}$ and $JSD_{Hierarchy}$ varied by model parameters $p$ and $k$ in Fig.~\ref{fig:compare_fig}a and the $JSD_{Radius}$ varied by model parameter $\gamma$ in Fig.~\ref{fig:compare_fig}b. For both periods, the overall fitness landscape does not change, while the optimal point does. In response to the COVID-19 pandemic, the likelihood of finding places $p$ decreases ($p_{pre}=0.820 > p_{post}=0.800$), indicating people prefer to stay in familiar places. In addition, the tendency to explore a high-level place decreases ($k_{pre}=2.075 > k_{post}=2.025$), and people become reluctant to travel far from their recreational homes ($\gamma_{pre}=1.325 < \gamma_{post}=1.375$). To check how much difference the parameter changes make, we simulated the model with the optimal parameters from the pre-COVID-19 period ($p=0.820$ and $k=2.075$) for the post-COVID-19 period with an assumption that user behaviors do not change. We checked that $JSD_{Entropy} \times JSD_{Hierarchy}$ of this model increases by 7\% compared to our optimal model. Similarly, if we simulate the model with the previous optimal $\gamma$ from the pre-COVID-19 period, $JSD_{Radius}$ increases by 16\%.

The differences in the estimated parameters between the two periods are not subtle, and the model's goodness of fit is sensitive to the parameter changes. In Fig.~\ref{fig:compare_fig}a, if $p$ increases or decreases by 0.02 from the estimated optimal point $p^*$, $JSD_{Entropy} \times JSD_{Hierarchy}$ increases by a factor of 1.058 and 1.080, respectively. If $k$ increases or decreases by 0.025 from the estimated optimal point $k^*$, $JSD_{Entropy} \times JSD_{Hierarchy}$ increases by a factor of 1.119 and 1.057, respectively. Similarly, if $\gamma$ increases or decreases by 0.025 from the estimated optimal point $\gamma^*$, $JSD_{Radius}$ increases by a factor of 1.022, 1.024, respectively. Also, we would like to note that the goodness of fit's standard deviation for ten repetitions is an order of magnitude smaller than the average value of goodness of fit, implying our simulation results are robust to random errors.

Intuitively, before the pandemic, people prefer to visit popular places as they have fewer restrictions on geographic distance. However, during the pandemic, people chose familiar (high $p$), relatively less popular (low $k$), and closer places (large $\gamma$) from their recreational homes. Low $k$ would reflect the behaviors of avoiding dense areas to prevent exposure to COVID-19, and large $\gamma$ would be associated with the tendency not to take public transportation. After the outbreak, the public transportation system usage in Seoul declined sharply: $-26.5$\% for the bus, $-27.5$\% for the subway in 2020 compared to the previous year, according to a report from the Korean Ministry of Land, Infrastructure, and Transport~\cite{kmit_stat2021}. Furthermore, these changes imply the effect of COVID-19 on urban inequality. As people avert crowded places but choose less popular places, the concentration of activities at the highest level is dissolved. However, the number of visitations at low hierarchy levels ($l\leq6$) also decreases (Supplementary Fig.~S9) because the probability of exploring places $p^i$ quickly converges to zero by iterations.

\section*{Discussion}

In this paper, we quantify the characteristics of recreational visitations with three factors: recreational  hierarchy, geographical distance, and attachment to a location. Leveraging our findings, we develop a model that successfully reconstructs and explains empirical patterns found in Seoul, Republic of Korea. We show that the COVID-19 pandemic led people to be less likely to visit different levels in the hierarchy. They prefer familiar, less popular, and closer locations. Furthermore, we suggest a possible mechanism to explain the worsening inequality with the model parameters $p$ and $k$ simultaneously.
 
Our study has several limitations. First, agents start from the empty reservation history and find a place by the model mechanisms. In reality, each individual could have past reservation records and find a place depending on the given history. Second, we use the geographic distance between cells, while the urban transportation systems can distort the distance. Third, our data are strongly correlated with mobility inflows. Therefore, changes in motel booking behaviors would explain changes in recreational activities. Considering diverse layers and their interactions can deeply enrich the understanding of urban activity. Fourth, our findings on behavioral changes could be the combination of voluntary willingness to avoid physical contact and public health guidelines such as social distancing policies, operating hours restrictions, and maximum occupancy restrictions. People might gather in less popular places to avoid waiting because many facilities could handle fewer people than before due to capacity limits. Additionally, they might prefer to gather in closer locations to return home without spending much time on public transportation. It would be interesting to decompose the effects of these factors with high-resolution data and advanced models. Fifth, our findings on the behavioral changes would be mainly led by low- and middle-income populations, the primary users of the platform we studied.

Despite these limitations, our study enhances the understanding of urban human activities and would help design effective public health policies considering individual movements around home areas. Practically, our model can be a simulation tool to prepare for unexpected future events that may affect human behaviors. With our model, academic and industry researchers can also tackle inequality issues stemming from behavioral changes across the urban hierarchy.

\section*{Materials and Methods}

\subsection*{Accommodation reservation data}

We sourced an accommodation reservation data set from Goodchoice Company LTD, a Korean platform that occupied 29\% of the online market share in 2020 (Wiseapp Report, \url{https://www.dailypop.kr/news/articleView.html?idxno=51946}, a news article written in Korean). The data contain anonymized customer-level reservation histories, spanning the period from January 2019 to November 2020, and geographic locations of 1,038 unique motels in Seoul, Republic of Korea. No demographic information is available.

\subsection*{Seoul mobility data}
Seoul mobility data was downloaded from the Seoul Open Data Plaza (\url{https://data.seoul.go.kr/dataVisual/seoul/seoulLivingMigration.do}, Accessed on December 3, 2021). The data contains mobility flows between administrative divisions (425 divisions in total) decomposed by gender, age, time of departure, and time of arrival, by aggregating the mobile phone signals from transceiver stations in Seoul. Also, for each individual, the data provide an estimated daytime residence (denoted as ``W'', mostly workplace), nighttime residence (denoted as ``H'', mostly home), and other classes (denoted as ``E''). With the classifications above, we can infer the context of urban mobility. For instance, a movement from a workplace to another area to enjoy recreational gathering is classified as the ``WE'' type. We use the mobility data spanning the period from January 2020 to October 2020 and focus on the mobility types with ``WE'', ``HE'', and ``EE'' to track non-routine mobility patterns for recreational gatherings.
 
\subsection*{Hotspot entropy and the radius of recreational activities}
Based on recreational hierarchy, we characterize location trajectories with two proposed measures: \textit{hotspot entropy} and \textit{radius of recreational activities}. First, the hotspot entropy is the Shannon entropy of hotspot levels in a trajectory. It is defined as 
\begin{equation}
    h = - \sum^{L}_{i=1}p_i \log {p_i},
\end{equation}
where $L$ is the total number of hotspot levels ($L=10$) and $p_i$ is the frequency of a hotspot level $i$ in the trajectory. For example in Fig.~\ref{fig:data_fig}a, $p$ is $[\frac{2}{3}, \frac{1}{6}, \frac{1}{6},  0, 0, 0, 0, 0, 0, 0]$ and hotspot entropy $h$ is $- \frac{2}{3}\log{\frac{2}{3}} - \frac{1}{6}\log{\frac{1}{6}} - \frac{1}{6}\log{\frac{1}{6}} = 0.867$.

Second, to quantify how far the places are in a trajectory, we define the radius of recreational activities. It is the variance of geographic distances from the most frequent cell (home cell) in a trajectory (similar to the radius of gyration) and is calculated as
\begin{equation}
    r = \sqrt{\frac{\sum_{i=1}^{N} d_{i,home}^2}{N}},
\end{equation}
where $N$ is the length of a trajectory, $d$ is the Haversine distance between the centers of the two cells, and $home$ is the \textit{recreational home} cell that is the most frequent cell in the trajectory. If there are multiple most frequent cells, we randomly pick one as the home cell.

\section*{Declarations}

\subsection*{Availability of data and materials}
Due to privacy concerns, the accommodation reservation data we used cannot be shared publicly. The Seoul mobility data is publicly available. 

\subsection*{Competing interests}
The authors have no competing interests.

\subsection*{Ethical approval statements}
This article does not contain any studies with human participants performed by any of the authors.

\subsection*{Informed consent statements}
This article does not contain any studies with human participants performed by any of the authors.

\subsection*{Funding}
This work was supported by the National Research Foundation of Korea (NRF) with grant number 2021R1F1A106303011. J.Y. was supported by the National Science Foundation with grant number 2133863.

\subsection*{Author's contribution}
All authors contributed to the work presented in this paper. J.Y. was involved in conceptualization, analysis, and writing, W.-S.J. and H.K. contributed to conceptualization and writing. All authors discussed the results and commented on the manuscript at all stages.

\subsection*{Acknowledgements}
We thank Goodchoice Company LTD. for making the accommodation reservation data available for this research. We thank I. Hong, O.-H. Kwon, D. Lee, and Y.-Y. Ahn for their helpful discussions.

\printbibliography{}

@article{bassolas2019hierarchical,
  title={Hierarchical organization of urban mobility and its connection with city livability},
  author={Bassolas, Aleix and Barbosa-Filho, Hugo and Dickinson, Brian and Dotiwalla, Xerxes and Eastham, Paul and Gallotti, Riccardo and Ghoshal, Gourab and Gipson, Bryant and Hazarie, Surendra A and Kautz, Henry and others},
  journal={Nature Communications},
  volume={10},
  number={1},
  pages={1--10},
  year={2019},
  publisher={Nature Publishing Group}
}

@article{chang2021mobility,
  title={Mobility network models of COVID-19 explain inequities and inform reopening},
  author={Chang, Serina and Pierson, Emma and Koh, Pang Wei and Gerardin, Jaline and Redbird, Beth and Grusky, David and Leskovec, Jure},
  journal={Nature},
  volume={589},
  number={7840},
  pages={82--87},
  year={2021},
  publisher={Nature Publishing Group}
}

@article{kang2020coronavirus,
  title={Coronavirus disease exposure and spread from nightclubs, South Korea},
  author={Kang, Cho Ryok and Lee, Jin Yong and Park, Yoojin and Huh, In Sil and Ham, Hyon Jeen and Han, Jin Kyeong and Kim, Jung Il and Na, Baeg Ju and COVID, Seoul Metropolitan Government and Team, Rapid Response},
  journal={Emerging Infectious Diseases},
  volume={26},
  number={10},
  pages={2499},
  year={2020},
  publisher={Centers for Disease Control and Prevention}
}

@article{roth2011structure,
  title={Structure of urban movements: Polycentric activity and entangled hierarchical flows},
  author={Roth, Camille and Kang, Soong Moon and Batty, Michael and Barth{\'e}lemy, Marc},
  journal={PLoS One},
  volume={6},
  number={1},
  pages={e15923},
  year={2011},
  publisher={Public Library of Science}
}

@article{louail2015uncovering,
  title={Uncovering the spatial structure of mobility networks},
  author={Louail, Thomas and Lenormand, Maxime and Picornell, Miguel and Cant{\'u}, Oliva Garc{\'\i}a and Herranz, Ricardo and Frias-Martinez, Enrique and Ramasco, Jos{\'e} J and Barth{\'e}lemy, Marc},
  journal={Nature Communications},
  volume={6},
  number={1},
  pages={1--8},
  year={2015},
  publisher={Nature Publishing Group}
}

@article{verbavatz2020growth,
  title={The growth equation of cities},
  author={Verbavatz, Vincent and Barth{\'e}lemy, Marc},
  journal={Nature},
  volume={587},
  number={7834},
  pages={397--401},
  year={2020},
  publisher={Nature Publishing Group}
}

@article{bettencourt2020demography,
  title={Demography and the emergence of universal patterns in urban systems},
  author={Bettencourt, Lu{\'\i}s MA and Z{\"u}nd, Daniel},
  journal={Nature Communications},
  volume={11},
  number={1},
  pages={1--9},
  year={2020},
  publisher={Nature Publishing Group}
}

@article{storz2015mobility,
  title={Mobility and innovation: A cross-country comparison in the video games industry},
  author={Storz, Cornelia and Riboldazzi, Federico and John, Moritz},
  journal={Research Policy},
  volume={44},
  number={1},
  pages={121--137},
  year={2015},
  publisher={Elsevier}
}

@article{park2019global,
  title={Global labor flow network reveals the hierarchical organization and dynamics of geo-industrial clusters},
  author={Park, Jaehyuk and Wood, Ian B and Jing, Elise and Nematzadeh, Azadeh and Ghosh, Souvik and Conover, Michael D and Ahn, Yong-Yeol},
  journal={Nature Communications},
  volume={10},
  number={1},
  pages={1--10},
  year={2019},
  publisher={Nature Publishing Group}
}

@book{batty2013new,
  title={The new science of cities},
  author={Batty, Michael},
  year={2013},
  publisher={MIT press}
}

@book{barthelemy2016structure,
  title={The structure and dynamics of cities},
  author={Barth{\'e}lemy, Marc},
  year={2016},
  publisher={Cambridge University Press}
}

@article{pan2013urban,
  title={Urban characteristics attributable to density-driven tie formation},
  author={Pan, Wei and Ghoshal, Gourab and Krumme, Coco and Cebrian, Manuel and Pentland, Alex},
  journal={Nature Communications},
  volume={4},
  number={1},
  pages={1--7},
  year={2013},
  publisher={Nature Publishing Group}
}

@article{nouvellet2021reduction,
  title={Reduction in mobility and COVID-19 transmission},
  author={Nouvellet, Pierre and Bhatia, Sangeeta and Cori, Anne and Ainslie, Kylie EC and Baguelin, Marc and Bhatt, Samir and Boonyasiri, Adhiratha and Brazeau, Nicholas F and Cattarino, Lorenzo and Cooper, Laura V and others},
  journal={Nature Communications},
  volume={12},
  number={1},
  pages={1--9},
  year={2021},
  publisher={Nature Publishing Group}
}

@incollection{batty2021london,
  title={London in lockdown: mobility in the pandemic city},
  author={Batty, Michael and Murcio, Roberto and Iacopini, Iacopo and Vanhoof, Maarten and Milton, Richard},
  booktitle={COVID-19 Pandemic, Geospatial Information, and Community Resilience},
  pages={229--244},
  year={2021},
  publisher={CRC Press}
}

@article{albert2000error,
  title={Error and attack tolerance of complex networks},
  author={Albert, R{\'e}ka and Jeong, Hawoong and Barab{\'a}si, Albert-L{\'a}szl{\'o}},
  journal={Nature},
  volume={406},
  number={6794},
  pages={378--382},
  year={2000},
  publisher={Nature Publishing Group}
}

@article{pastor2003epidemics,
  title={Epidemics and immunization in scale-free networks},
  author={Pastor-Satorras, Romualdo and Vespignani, Alessandro and others},
  journal={Handbook of Graphs and Networks},
  year={2003},
  publisher={Wiley Online Library}
}

@article{moro2021mobility,
  title={Mobility patterns are associated with experienced income segregation in large US cities},
  author={Moro, Esteban and Calacci, Dan and Dong, Xiaowen and Pentland, Alex},
  journal={Nature Communications},
  volume={12},
  number={1},
  pages={1--10},
  year={2021},
  publisher={Nature Publishing Group}
}

@article{van2020covid,
  title={COVID-19 exacerbating inequalities in the US},
  author={Van Dorn, Aaron and Cooney, Rebecca E and Sabin, Miriam L},
  journal={Lancet},
  volume={395},
  number={10232},
  pages={1243},
  year={2020},
  publisher={Elsevier}
}

@article{ahmed2020inequality,
  title={Why inequality could spread COVID-19},
  author={Ahmed, Faheem and Ahmed, Na'eem and Pissarides, Christopher and Stiglitz, Joseph},
  journal={The Lancet Public Health},
  volume={5},
  number={5},
  pages={e240},
  year={2020},
  publisher={Elsevier}
}

@article{yu2021racial,
  title={Racial residential segregation and economic disparity jointly exacerbate COVID-19 fatality in large American cities},
  author={Yu, Qinggang and Salvador, Cristina E and Melani, Irene and Berg, Martha K and Neblett, Enrique W and Kitayama, Shinobu},
  journal={Annals of the New York Academy of Sciences},
  year={2021},
  publisher={Wiley-Blackwell}
}

@article{gould1994spatial,
  title={Spatial structures and scientific paradoxes in the AIDS pandemic},
  author={Gould, Peter and Wallace, Rodrick},
  journal={Geografiska Annaler: Series B, Human Geography},
  volume={76},
  number={2},
  pages={105--116},
  year={1994},
  publisher={Taylor \& Francis}
}

@article{kandampully2000customer,
  title={Customer loyalty in the hotel industry: The role of customer satisfaction and image},
  author={Kandampully, Jay and Suhartanto, Dwi},
  journal={International Journal of Contemporary Hospitality Management},
  year={2000},
  publisher={MCB UP Ltd}
}

@article{bowen2001relationship,
  title={The relationship between customer loyalty and customer satisfaction},
  author={Bowen, John T and Chen, Shiang-Lih},
  journal={International Journal of Contemporary Hospitality Management},
  year={2001},
  publisher={MCB UP Ltd}
}

@misc{ googles2, title={S2 geometry}, url={https://s2geometry.io/}, journal={S2Geometry}, author={ Veach, Eric and Rosenstock, Jesse and Eagle, Eric and Snedegar, Robert and Basch, Julien}, year={2017}}

@article{wu2018travel,
  title={Travel time estimation using spatio-temporal index based on Cassandra},
  author={Wu, Zheng and Li, Chengming and Wu, Yinghao and Xiao, Fei and Zhu, Lining and Shen, Jianming},
  journal={ISPRS Annals of Photogrammetry, Remote Sensing \& Spatial Information Sciences},
  volume={4},
  number={4},
  year={2018}
}

@article{narula2020policy,
  title={Policy opportunities and challenges from the COVID-19 pandemic for economies with large informal sectors},
  author={Narula, Rajneesh},
  journal={Journal of International Business Policy},
  volume={3},
  number={3},
  pages={302--310},
  year={2020},
  publisher={Springer}
}

@article{song2010modelling,
  title={Modelling the scaling properties of human mobility},
  author={Song, Chaoming and Koren, Tal and Wang, Pu and Barab{\'a}si, Albert-L{\'a}szl{\'o}},
  journal={Nature Physics},
  volume={6},
  number={10},
  pages={818--823},
  year={2010},
  publisher={Nature Publishing Group}
}

@article{schlapfer2021universal,
  title={The universal visitation law of human mobility},
  author={Schl{\"a}pfer, Markus and Dong, Lei and O’Keeffe, Kevin and Santi, Paolo and Szell, Michael and Salat, Hadrien and Anklesaria, Samuel and Vazifeh, Mohammad and Ratti, Carlo and West, Geoffrey B},
  journal={Nature},
  volume={593},
  number={7860},
  pages={522--527},
  year={2021},
  publisher={Nature Publishing Group}
}

@article{kim2018qualitative,
  title={A qualitative approach to automated motels: a rising issue in South Korea},
  author={Kim, Hyojin and Kim, Byung-Gook},
  journal={International Journal of Contemporary Hospitality Management},
  year={2018},
  publisher={Emerald Publishing Limited}
}

@book{lashley2016routledge,
  title={The Routledge handbook of hospitality studies},
  author={Lashley, Conrad},
  year={2016},
  chapter={28},
  publisher={Taylor \& Francis}
}

@article{Hu2016,
 author  = {Jeongyeon Hu},
 date    = {2016-04-02},
 title   = {Trend of using motels for the 2030 generation},
 journal = {The JoongAng},
 url     = {https://www.joongang.co.kr/article/19826193#home},
 urldate = {2016-04-02}
}

@misc{accomodationstat2021,
  title = {Domestic \& Global Accommodation Platform Analysis},
  howpublished = {\url{https://market.dighty.com/trendreport/?q=YToxOntzOjEyOiJrZXl3b3JkX3R5cGUiO3M6MzoiYWxsIjt9&bmode=view&idx=8435265&t=board}},
  note = {Accessed: 2022-08-01}
}

@article{zhang2022unequal,
  title={The unequal effect of the COVID-19 pandemic on the labour market and income inequality in China: A multisectoral CGE model analysis coupled with a micro-simulation approach},
  author={Zhang, Qi and Zhang, Xinxin and Cui, Qi and Cao, Weining and He, Ling and Zhou, Yexin and Li, Xiaofan and Fan, Yunpeng},
  journal={International Journal of Environmental Research and Public Health},
  volume={19},
  number={3},
  pages={1320},
  year={2022},
  publisher={MDPI}
}

@article{hevia2022covid,
  title={Covid-19 in unequal societies},
  author={Hevia, Constantino and Macera, Manuel and Neumeyer, Pablo Andr{\'e}s},
  journal={Journal of Economic Dynamics and Control},
  pages={104328},
  year={2022},
  publisher={Elsevier}
}

@article{belot2021unequal,
  title={Unequal consequences of Covid 19: representative evidence from six countries},
  author={Belot, Mich{\`e}le and Choi, Syngjoo and Tripodi, Egon and Broek-Altenburg, Eline van den and Jamison, Julian C and Papageorge, Nicholas W},
  journal={Review of Economics of the Household},
  volume={19},
  number={3},
  pages={769--783},
  year={2021},
  publisher={Springer}

}

@misc{kmit_stat2021,
  title = {Changes in public transportation usage due to COVID-19},
  howpublished = {\url{http://www.molit.go.kr/USR/NEWS/m_71/dtl.jsp?lcmspage=1&id=95085321}},
  note = {Accessed: 2022-08-05}
}

\begin{center}
\textbf{\Huge Supplementary Information : COVID-19 confines recreational gatherings in Seoul to familiar, less crowded, and neighboring urban areas}
\end{center}
\setcounter{equation}{0}
\setcounter{figure}{0}
\setcounter{table}{0}
\makeatletter
\renewcommand{\theequation}{S\arabic{equation}}
\renewcommand{\thefigure}{S\arabic{figure}}

\begin{figure}[ht]
\centering
\includegraphics[width=0.5\linewidth]{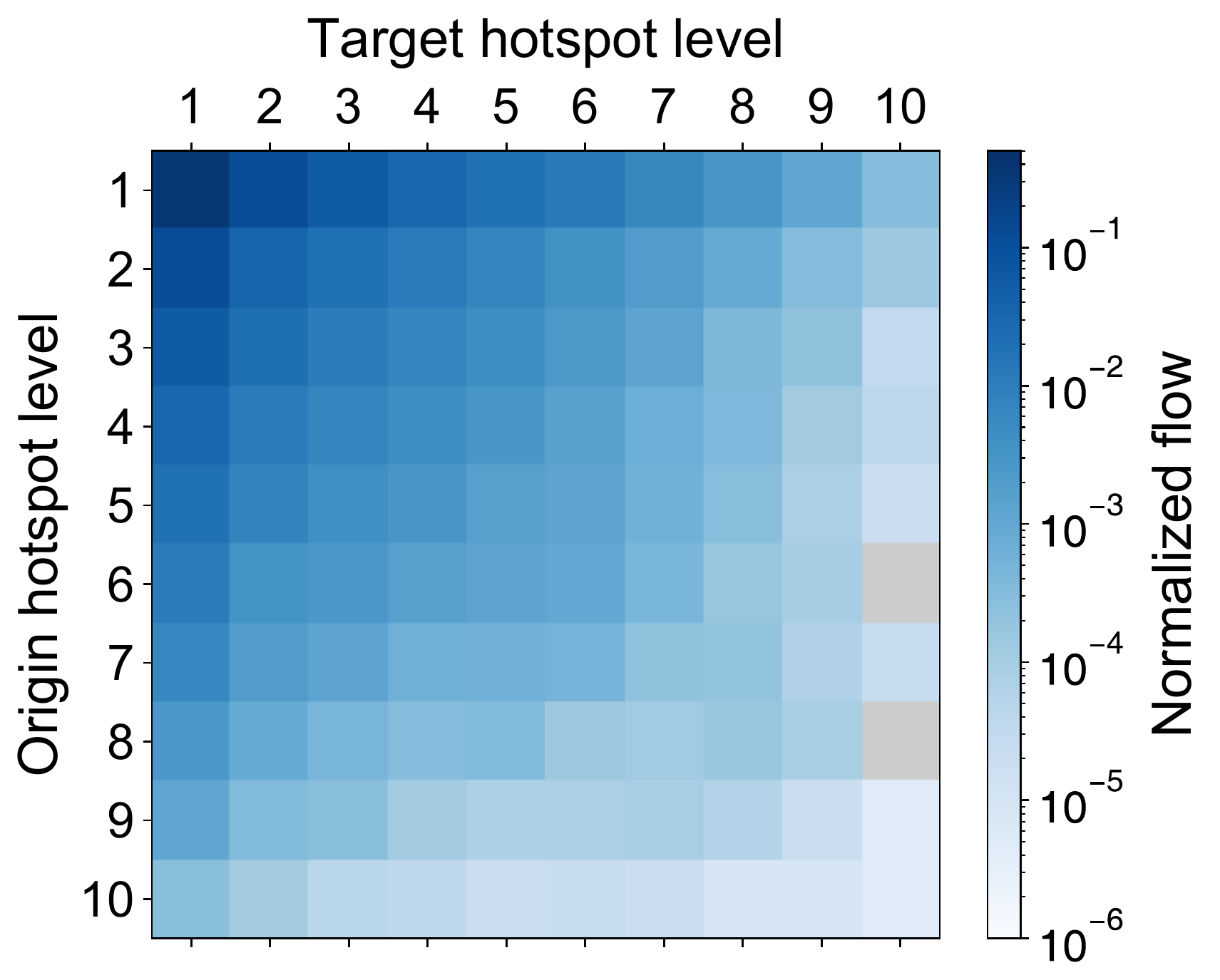}
\caption{\textbf{The flow matrix $T^{data}$ for the post-COVID-19 period.} As same as Fig.~1 in the main report, we exclude the trips within the same cell. There are no transition records for the gray cells.}
\label{suppfig:flow_matrix_after}
\end{figure}

\begin{figure}[ht]
\centering
\includegraphics[width=\linewidth]{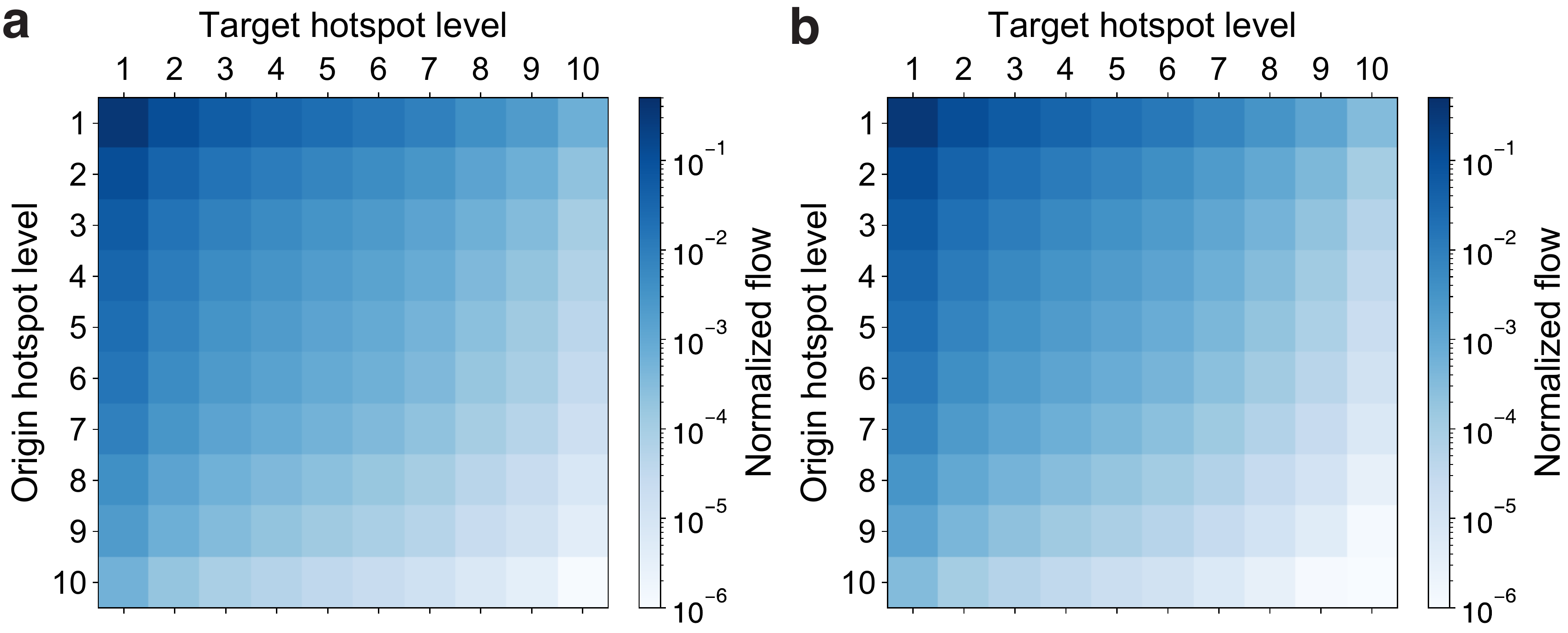}
\caption{\textbf{The flow matrix of the null model, $T^{null}$, for the (a) pre-COVID-19 period and the (b) post-COVID-19 period.}}
\label{suppfig:null_flow_matrix}
\end{figure}

\begin{figure}[ht]
\centering
\includegraphics[width=\linewidth]{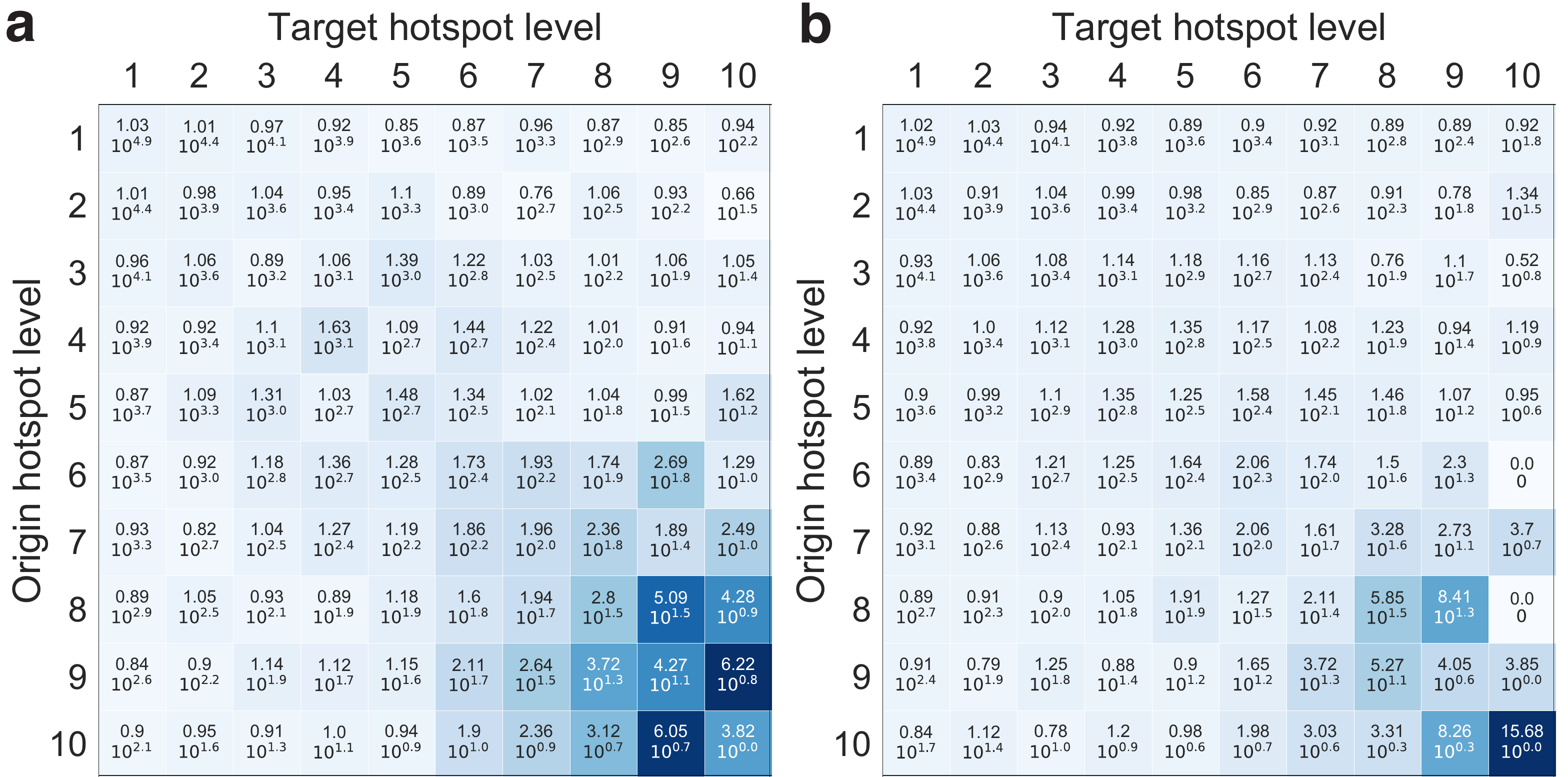}
\caption{ \textbf{The ratio matrix $T^{data}/T^{null}$ in the (a) pre-COVID-19 period and the (b) post-COVID-19 period.} In each cell, upper annotated number is $T_{data}/T_{null}$ and below annotated number is the total number of transitions in the data.}
\label{suppfig:ratio_matrix}
\end{figure}

\begin{figure}[ht]
\centering
\includegraphics[width=\linewidth]{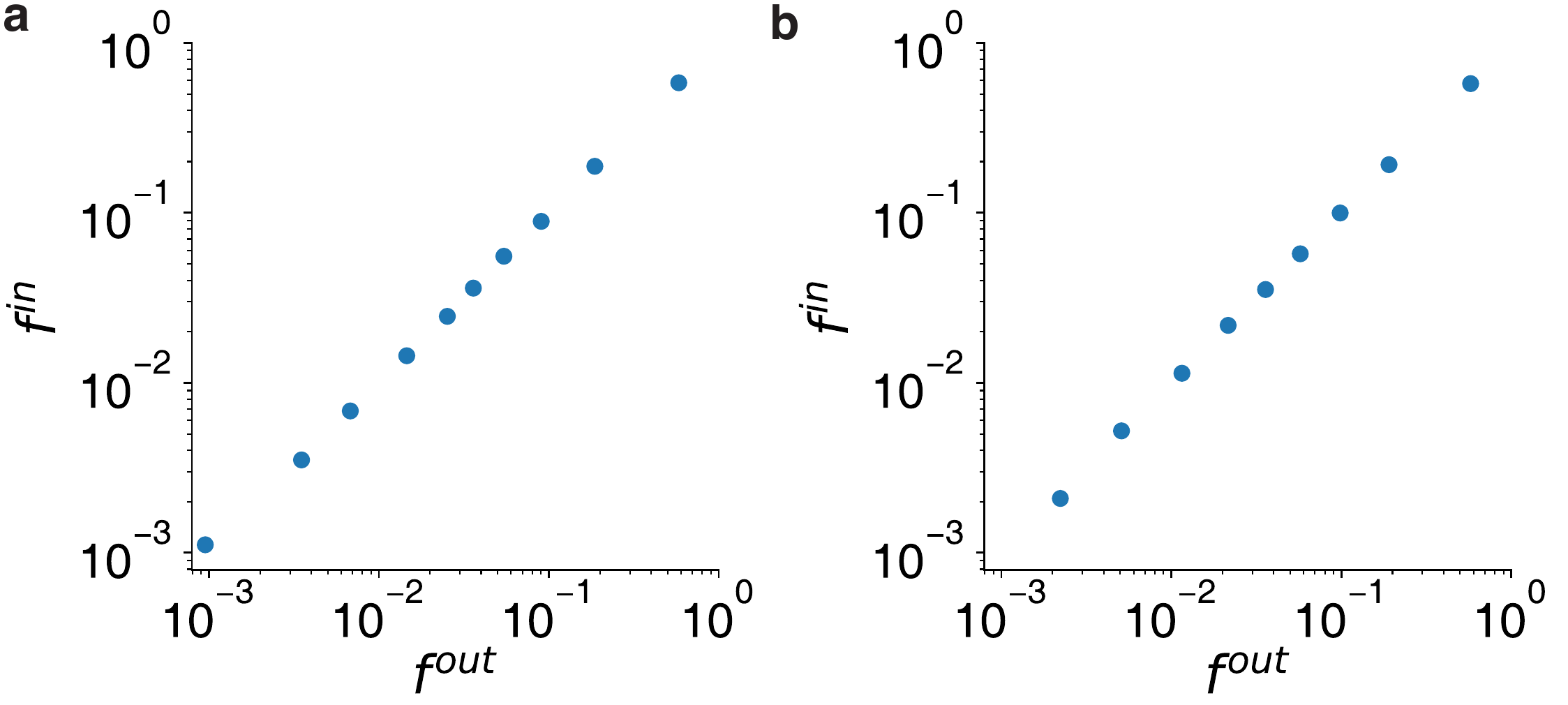}
\caption{\textbf{Transition outflows and inflows in (a) the pre-COVID-19 period and (b) the post-COVID-19 period.} The infow is defined as $f^{in}_i = \sum_{k=1}^L T_{ki}$ and outflow is defined as $f^{out}_i=\sum_{k=1}^L T_{ik}$ where $L$ is the total number of hotspot level. $f^{in}$ and $f^{out}$ is almost symmetry for both periods ($R^2 > 0.999$).
}
\label{suppfig:in_and_out}
\end{figure}

\begin{figure}[ht]
\centering
\includegraphics[width=0.7\linewidth]{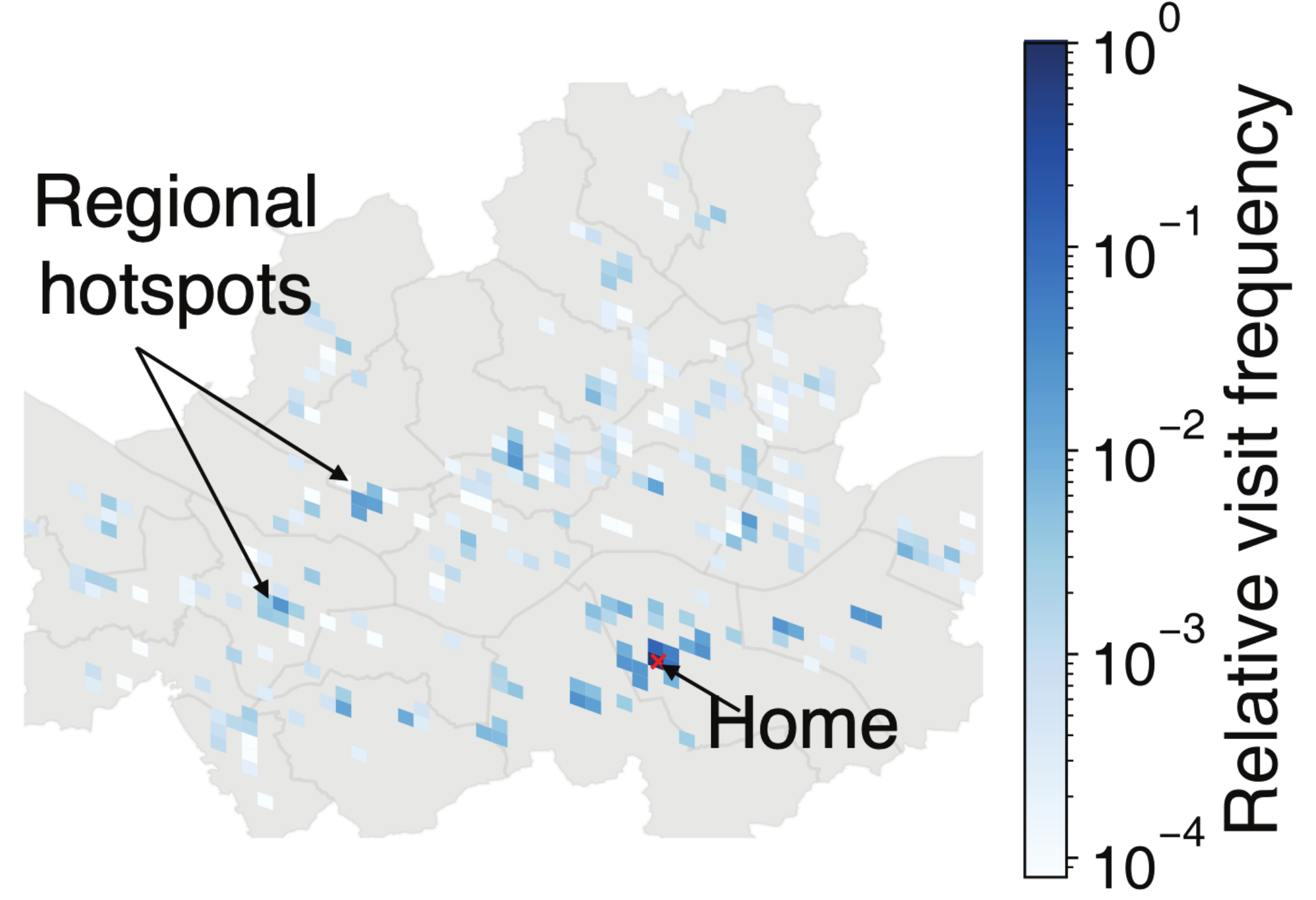}
\caption{\textbf{Urban hierarchy leads people to move farther than expected.} We collect the trajectories of which recreational home cells are near the Gangnam area (red cross, hotspot level 1), which is one of popular regions in Seoul. Relative visit frequency decays with the geographic distance from the home cell, but the geographic distance cannot explain the pattern near regional hotspots.
}
\label{suppfig:regional_hotspot}
\end{figure}

\begin{figure}[ht]
\centering
\includegraphics[width=\linewidth]{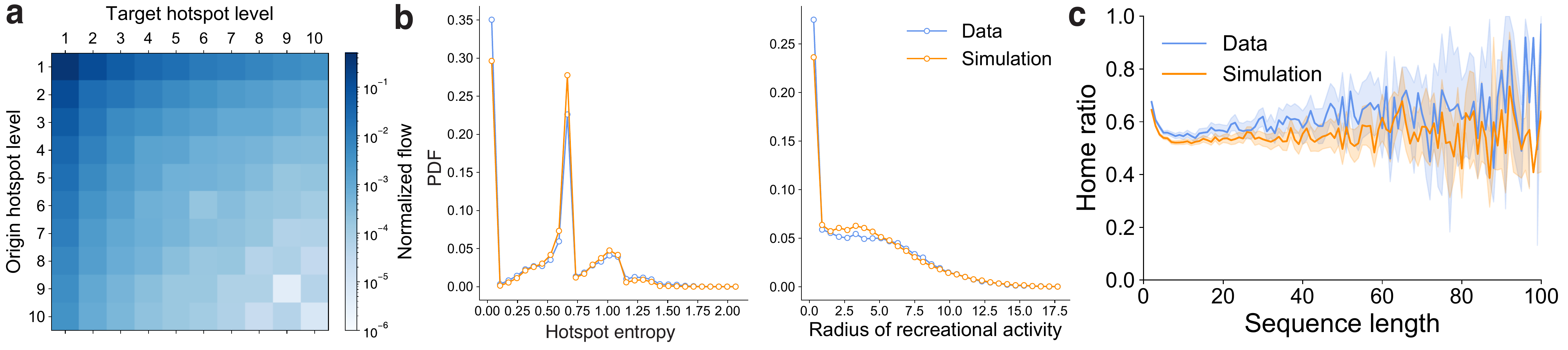}
\caption{\textbf{The best model result for the post-COIVD-19 period} \textbf{(a)} The flow matrix $T^{model}$ of the best model, $d_T = ||T^{model} - T^{data}||_F$ is 0.02. \textbf{(b) The hotspot entropy distribution $p_h$ (left) and the distribution of the radius of recreational activities $p_r$ (right).} Blue lines are the empirical distributions, and orange lines are the simulation results. \textbf{(c) Home ratios by the length of trajectory.}
}
\label{suppfig:best_after}
\end{figure}

\begin{figure}[ht]
\centering
\includegraphics[width=\linewidth]{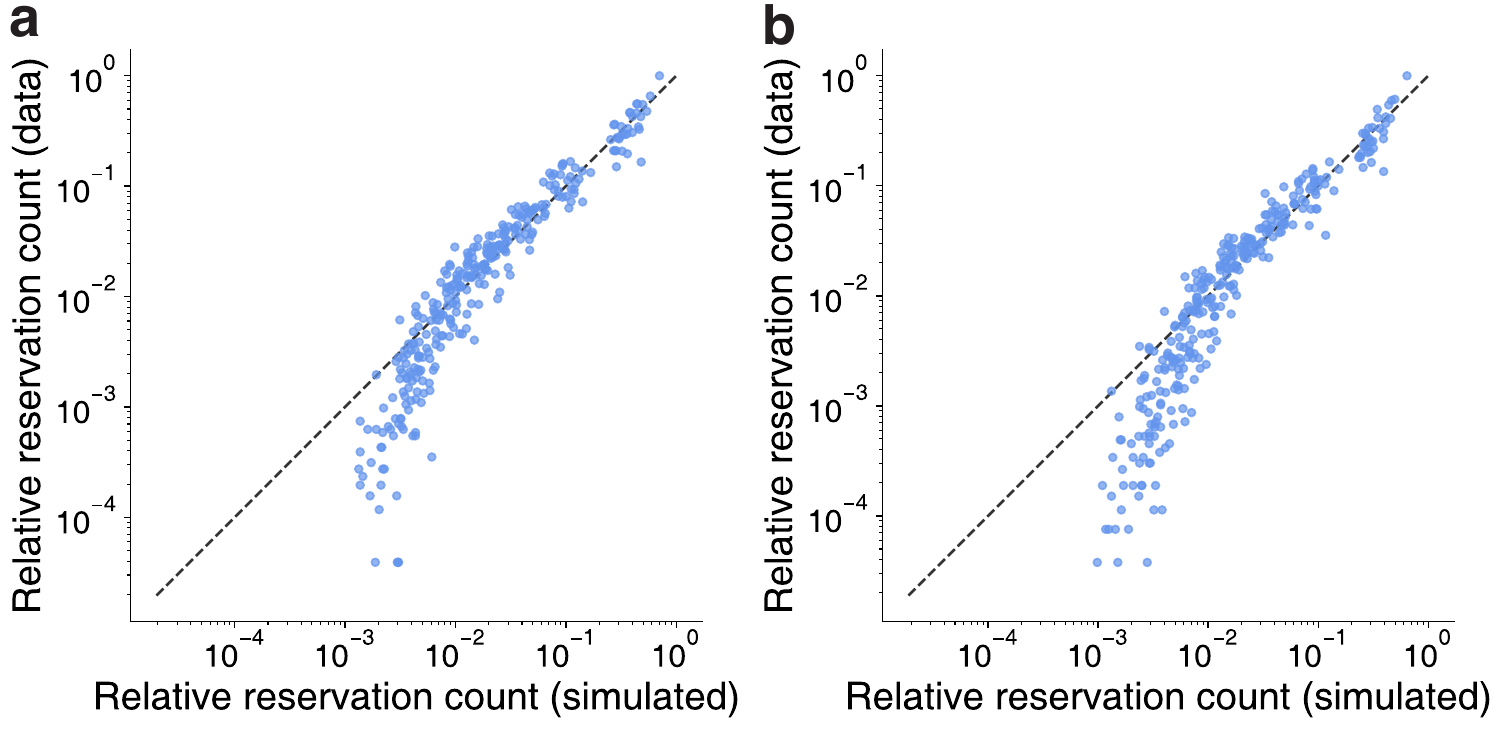}
\caption{\textbf{Cell-level comparisons between actual and simulated reservation counts for the (a) pre-COVID-19 period and the (b) post-COVID-19 period.}
Overall, our model simulates the cell-level reservation count well for both periods. For the data privacy concern, we normalize the reservation count with the maximum reservation count of the actual data set.
}
\label{suppfig:prediciton}
\end{figure}

\begin{figure}[ht]
\centering
\includegraphics[width=\linewidth]{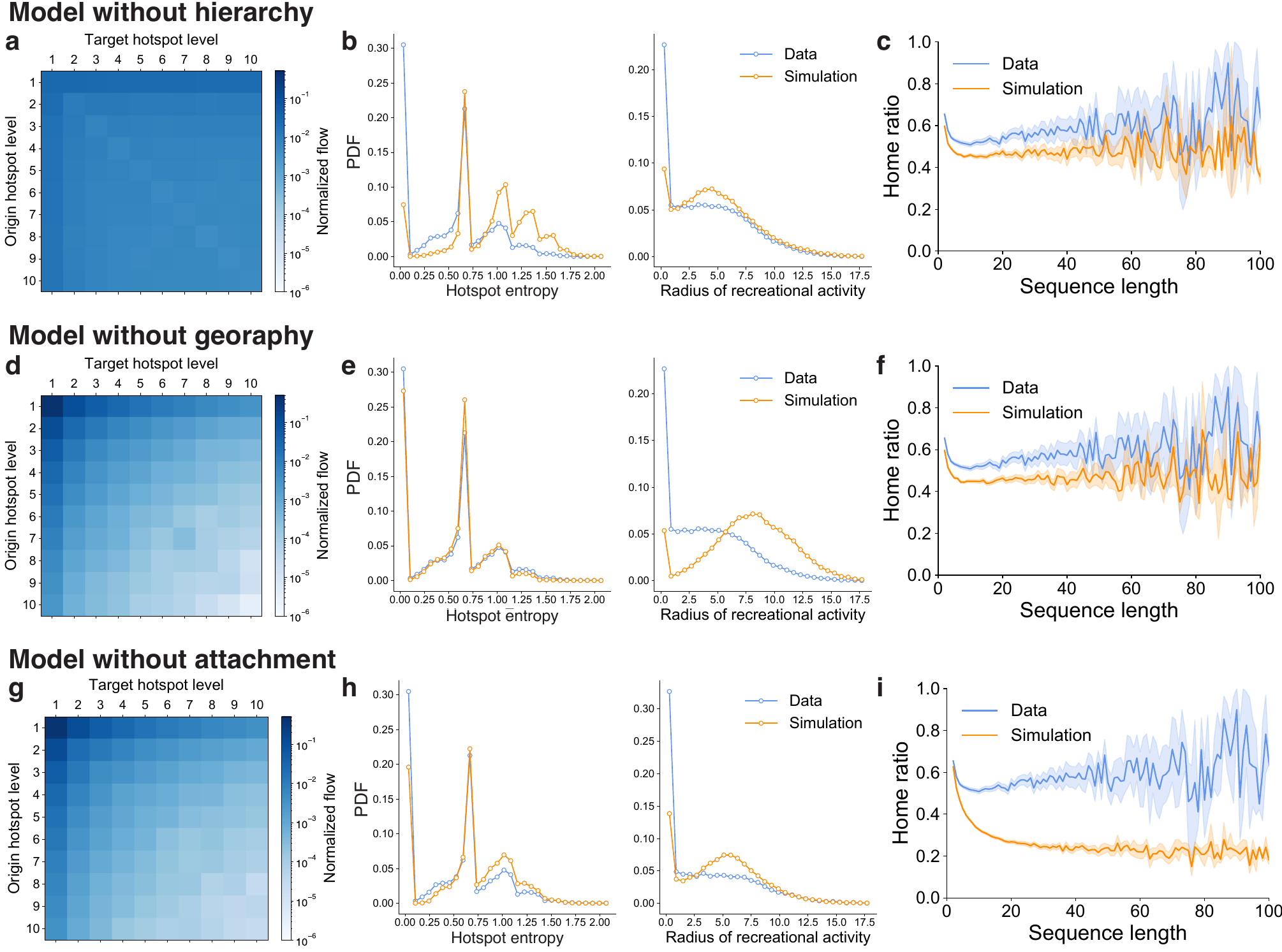}
\caption{\textbf{Variant model results for the pre-COIVD-19 period. (a)-(c) The model without urban hierarchy. (d)-(f) The model without geography. (g)-(h) The model without attachment to a location}. The figures in the first column are the flow matrices from each model $T^{model}$. The figures in the second column are the hotspot entropy distributions $p_h^{model}$ (left) and the radius of recreational activities distributions $p_r^{model}$ (right) from each model. Blue lines are the empirical distributions, and orange lines are the simulation results. Lastly, the figures in the third column are home ratios by sequence length. The overall pattern is similar to the pattern in the post-COVID-19 period. 
}
\label{suppfig:variant}
\end{figure}

\begin{figure}[ht]
\centering
\includegraphics[width=0.6\linewidth]{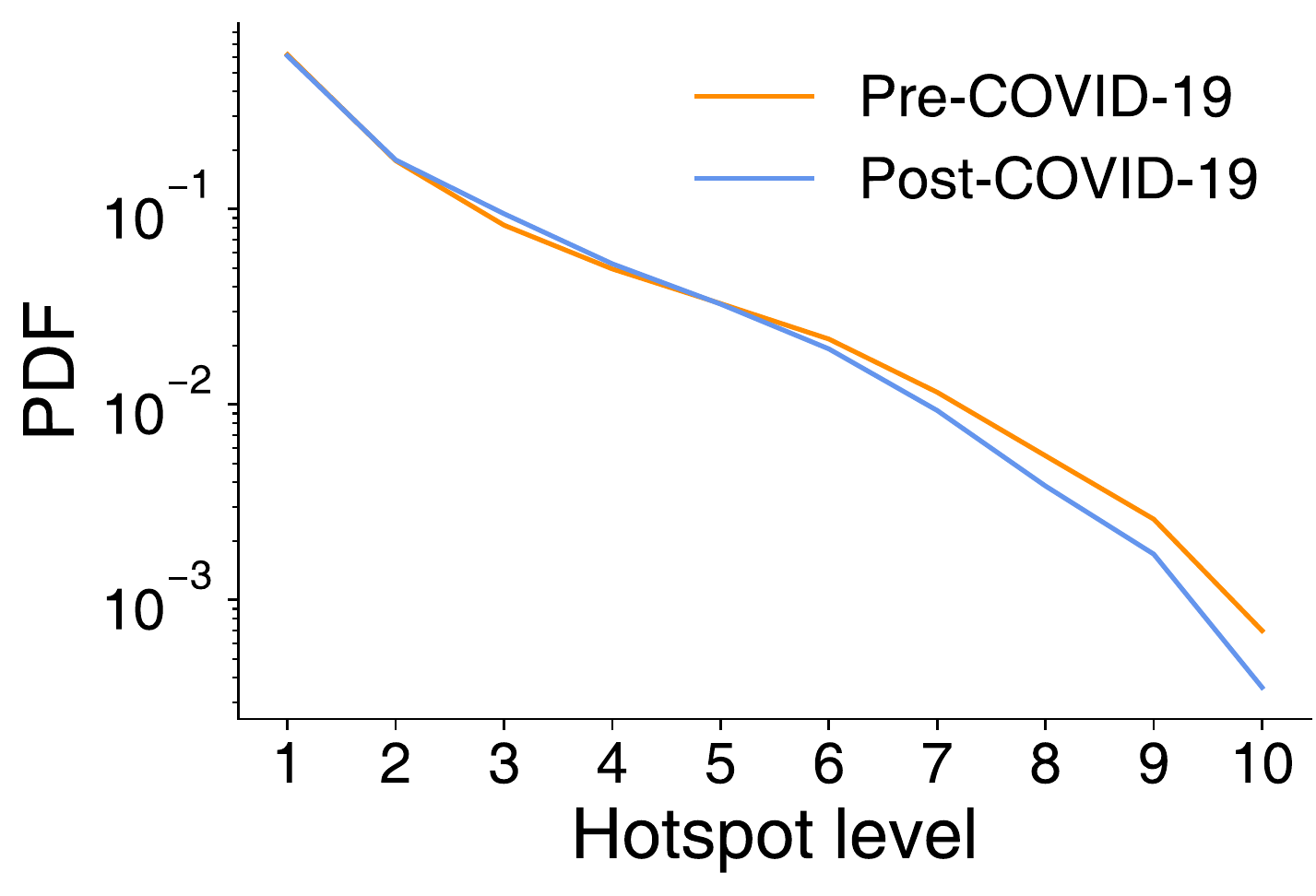}
\caption{\textbf{The simulated reservation count distribution by the hotspot level $p_{\ell}$.} The model successfully explains the decentralization of human urban activities and the worsening inequality of the urban areas in Seoul.
}
\label{suppfig:model_equality}
\end{figure}

\end{document}